\documentclass[modern]{aastex62}

\newcommand{\nicer}{\textit{NICER}}

\newcommand{\psr}{{PSR~J1231$-$1411}}

\newcommand{\Chisqred}{$\chi^2_\nu$}

\newcommand{\ee}[1]{\mbox{$10^{#1}$}}
\newcommand{\tee}[1]{\mbox{$\times 10^{#1}$}}
\newcommand{\percmsq}{\mbox{$\,{\rm cm^{-2}}$}}
\newcommand{\ud}[2]{\mbox{$^{+ #1}_{- #2}$}}
\newcommand{\ppm}{\mbox{$\pm$}}
\newcommand{\persec}{\mbox{$\,{\rm s^{-1}}$}}
\newcommand{\km}{\hbox{$\,{\rm km}$}}
\newcommand{\msun}{\mbox{$\,M_\odot$}}
\newcommand{\cgsflux}{\mbox{$\,{\rm erg\,\percmsq\,\persec}$}}

\shorttitle{X-ray Pulsations from PSR J1231$-$1411}
\shortauthors{Ray et al.}

\begin{document}

\title{Discovery of Soft X-ray Pulsations from PSR J1231$-$1411 using NICER}

\correspondingauthor{Paul S. Ray}
\email{paul.ray@nrl.navy.mil}

\author[0000-0002-5297-5278]{Paul S. Ray}
\affil{U.S. Naval Research Laboratory, Washington, DC 20375-5352, USA}
\author[0000-0002-6449-106X]{Sebastien Guillot}
\affil{IRAP, CNRS, 9 avenue du Colonel Roche, BP 44346, F-31028 Toulouse Cedex 4, France}
\affil{Universit\'{e} de Toulouse, CNES, UPS-OMP, F-31028 Toulouse, France.}
\author[0000-0001-5799-9714]{Scott M.~Ransom}
\affil{National Radio Astronomy Observatory, 520 Edgemont Road, Charlottesville, VA 22903, USA}
\author{Matthew Kerr}
\affil{U.S. Naval Research Laboratory, Washington, DC 20375-5352 USA}
\author[0000-0002-9870-2742]{Slavko Bogdanov}
\affil{Columbia Astrophysics Laboratory, Columbia University, 550 West 120th Street, New York, NY 10027, USA}
\author{Alice K. Harding}
\author[0000-0002-4013-5650]{Michael T. Wolff}
\affil{U.S. Naval Research Laboratory, Washington, DC 20375-5352, USA}
\author[0000-0002-0380-0041]{Christian Malacaria}
\affiliation{NASA Marshall Space Flight Center, NSSTC, 320 Sparkman Drive, Huntsville, AL 35805, USA}\thanks{NASA Postdoctoral Fellow}
\affiliation{Universities Space Research Association, NSSTC, 320 Sparkman Drive, Huntsville, AL 35805, USA}
\author{Keith C. Gendreau}
\affil{NASA's Goddard Space Flight Center, Greenbelt, MD, USA}
\author{Zaven Arzoumanian}
\affil{NASA's Goddard Space Flight Center, Greenbelt, MD, USA}
\author{Craig Markwardt}
\affil{NASA's Goddard Space Flight Center, Greenbelt, MD, USA}
\author{Yang Soong}
\affil{NASA's Goddard Space Flight Center, Greenbelt, MD, USA}
\affil{Universities Space Research Association, Columbia, MD 21044, USA}
\author{John P. Doty}
\affil{Noqsi Aerospace Ltd, 15 Blanchard Avenue, Billerica MA 01821, USA}
\noaffiliation



\begin{abstract}
We report the discovery of soft X-ray pulsations from the nearby millisecond pulsar PSR J1231$-$1411 using \nicer. The pulsed emission is characterized by a broad and asymmetric main pulse and a much fainter secondary interpulse, with a total pulsed count rate of 0.055 c s$^{-1}$ in the 0.35--1.5 keV band.  We analyzed \textit{Fermi} LAT data to update the pulse timing
model covering 10 years of data and used that model to coherently combine \nicer{} data over a year of observations. Spectral modeling suggests that the flux is dominated by 
thermal emission from a hot spot (or spots) on the neutron star surface. The phase relationship between the X-ray pulse and the radio and $\gamma$ rays provides insight into the geometry of the system.
\end{abstract}




\section{Introduction} \label{sec:intro}
Millisecond pulsars (MSPs) are an old (Gyr) population of neutron stars that are characterized by rapid spins ($P\lesssim 25$ ms) and exceptional rotational stability; they are the expected evolutionary outcomes of spin-up by accretion in X-ray binaries (\citealt{1982Natur.300..728A} for the seminal work; and \citealt{tauris06} for a recent review).

With a small number of notable exceptions, X-ray pulsations from rotation-powered MSPs typically have soft, blackbody-like spectra due to the cooling surface and/or hot spot(s) \citep[e.g.,][]{2006ApJ...638..951Z,2006ApJ...646.1104B}, with the bulk of photon flux detected below $\sim$2 keV and luminosities in the range $10^{29}-10^{31}$ erg s$^{-1}$.  This emission is commonly attributed to
return currents along open field lines heating the magnetic polar caps to temperatures $T_{\rm eff}\sim 10^6$ K \citep{harding01,harding02}. The X-ray pulsations tend to be broad but can have moderately high fractional pulsed amplitudes \citep[$\sim$30--70\%; see][]{2009ApJ...703.1557B,2013ApJ...762...96B}, implying an anisotropic emission pattern from the surface, such as may arise due to a light element neutron star atmosphere \citep[e.g.,][]{1987ApJ...313..718R,1996A&A...315..141Z}.

The \textit{Neutron Star Interior Composition Explorer} (\nicer; \citealt{NICERNatAs}) has been operating as an attached 
payload on the \textit{International Space Station} (ISS) since 2017 June. One of its primary science goals is to search for 
X-ray pulsations from a large number of known and candidate neutron star systems to reveal information on their energetics, 
evolution, and emission mechanisms. Another key motivation is to find rotation-powered millisecond pulsars that exhibit strong thermal pulsations
from surface hot spots. Careful modeling of the energy-dependent pulse profiles of MSPs can provide precise constraints on the mass 
and radius of the neutron star (\citealt{2007ApJ...670..668B}; see also \citealt{WattsRMP} for an overview). Prior to the \nicer\ launch, the most promising such pulsars known were
PSR J0437$-$4715 and J0030$+$0451.  In addition, the 3.68 ms pulsar J1231$-$1411 was known to emit thermal X-rays \citep{rrc+11},
but had never been observed with an instrument and mode capable of testing whether these X-rays were pulsed. Consequently,
this pulsar was the highest priority target for the \nicer\  Pulsation Search and Multiwavelength Coordination working group \citep{rag17}.

\psr\ was discovered in one of the first radio pulsation searches that targeted \textit{Fermi} Large Area Telescope (LAT) $\gamma$-ray sources that were unassociated with any probable counterpart \citep{rrc+11} at other wavelengths, a technique that turned out to be exceptionally successful \citep{RayPSC}, with at least 87 MSPs discovered to date. The pulsar is in a 1.86 day orbit about a white dwarf companion with minimum mass 0.19~$M_\sun$ and is nearby, with a dispersion measure distance of only 420 pc \citep{YMW16}. It is the brightest MSP in the $\gamma$-ray band with a flux $> 100$ MeV of 9.2(4) $\times 10^{-8}$ ph cm$^{-2}$ s$^{-1}$ ($1.03(3) \times 10^{-10}$ erg  cm$^{-2}$ s$^{-1}$; \citealt{2PC})\footnote{Here, and elsewhere in this paper, the number in parentheses is the uncertainty in the last digit.}.
Assuming a neutron star moment of inertia of $1\times 10^{45}$ g cm$^{2}$, the Shklovskii-corrected \citep{1970SvA....13..562S} spindown luminosity is $5\times 10^{33}$ erg s$^{-1}$, yielding a $\gamma$-ray efficiency ($L_\gamma/\dot{E}$) of 46\% \citep{2PC}. 


In this paper, we describe a deep \nicer\ observation of \psr\ and the analysis that resulted in the discovery of X-ray pulsations from this system.

\section{Observations}
\label{sec:obs}

\nicer's \citep{2016SPIEGendreau} X-ray Timing Instrument (XTI) is an array of 56 co-aligned X-ray optics
that concentrate X-rays in the 0.2--12 keV band onto an array of 56 single-pixel silicon drift detectors (52 currently functioning on orbit). Each optic is paired with a detector and associated
readout electronics, called Focal Plane Modules (FPMs).
The peak collecting area of the XTI is 1900 cm$^{2}$ at \replaced{1}{1.5} keV. All photons are individually time tagged with an achievable accuracy 
relative to GPS time of better than 100 ns \citep{2016SPIELaMarr,2016SPIEPrigozhin}.

\nicer\ observations are made up of short dwells that are a fraction of the \added{92-min} ISS orbit in duration (typically hundreds to $\sim$2500 
seconds).  All dwells from a given UTC day are grouped into a single ObsID for pipeline processing and delivery to the HEASARC 
archive. We collected data from 13 ObsIDs during the commissioning phase (prior to 2017 July 13) and 312 ObsIDs during the 
science operations phase up through 2018 July 26, for a total raw observing time of 1254.5 ks.

Event energies are defined by the \texttt{PI} column (the {\it Pulse Invariant}, in units of 10 eV) in the science data, which is computed from the raw pulse height by the 
\nicer\ data pipeline (version \texttt{10-master\_20180226} and CALDB \texttt{xti20180226}). 
In all of our event data, we filtered out events flagged as non-photon \replaced{events}{triggers} (see \citealt{2016SPIEPrigozhin}, for a detailed description of the detector system).
The detector electronics process pulses in parallel by two analysis chains, one with a slow shaping time (465 ns; optimized for precise 
energy measurements) and one with a fast shaping time (84 ns; optimized for precise time measurement). Each chain that triggers produces 
its own pulse height measurement. We accept only events where at least the slow channel is triggered and so our energy 
measurements are always based on the slow pulse height. The timing comes from the fast chain, unless it did not trigger, in which case
the slow chain is used. Because of the longer peaking time, there is a systematic offset in the time stamps of events that trigger 
only the slow chain. This fine clock correction is applied by the \nicer\ pipeline. 
The fast chain triggers for the majority of events above 1 keV \citep{2016SPIELaMarr}.
When both chains are triggered, we remove events where 
$\mathtt{PI\_RATIO} \equiv \mathtt{PI}/\mathtt{PI\_FAST} > 1.1 + 120/\mathtt{PI}$. This cut excludes events that occur far from the center of the detector, which are most likely to be particle events rather
than source photons.

We analyzed our data using HEASoft 6.25\footnote{Available at \url{https://heasarc.nasa.gov/lheasoft/}} and NICERDAS 2018-10-07\_V005. We initially selected good time intervals using \texttt{nimaketime} with the following screening criteria: 1) ISS outside the \nicer-specific
South Atlantic Anomaly (SAA) boundary; 2) \nicer\ in tracking mode with pointing direction $<0.015$\degr\ from the source direction with at least 38 detectors enabled; 3) source elevation $>20$\degr\ above the Earth limb;  4) source direction at least 30\degr\ from the bright Earth; and 5) magnetic cutoff rigidity $> 1.5$ GeV/$c$.  Applying these cuts resulted in a selection of 985 ks of filtered data. 

We have found that in some cases, the background (particularly at soft energies) is dominated by a few `hot' detectors, with some detectors much more likely to be affected by optical loading. In our analysis, we always excluded detector IDs 14, 34, and 54 for this reason.

Although the basic good time cuts exclude the very high background region
of the SAA, the ISS orbit also traverses regions of high latitude, referred to as the `polar horns', where the magnetic cutoff rigidity 
gets very low and particle backgrounds can be high and variable. We filter the worst parts 
of these regions by selecting a \deleted{low} minimum cutoff 
rigidity value (as described above). However, this cut does not fully remove all high background intervals in the data. Increasing the minimum rigidity will exclude more high background regions,
but would also exclude a substantial amount of time
where the background is low. Since the count rate from the pulsar is low ($\sim 0.2$ s$^{-1}$) and constant, we
filtered high background intervals using a count rate cut, which preserves the low background time in the polar horns. To accomplish this, we made a 16-second binned light curve 
of the 0.3--8 keV events (after all filtering described above).  We then filtered out all event 
data in bins where the count rate exceeded 2.5 s$^{-1}$.
This cut reduced our processed good time to 916 ks, which formed the basis for our analyses.


\section{Timing}

\subsection{Fermi LAT Timing}

We started from the pulsar timing model published with the \textit{Fermi} Second Pulsar Catalog
(2PC; \citealt{2PC}).  This timing model showed significant drift when extrapolated over the 5 
years since that publication, so we updated it using \textit{Fermi} LAT data, as described below.

Using the Pass 8 R2 data set \citep{P8}, we extracted LAT `Source class' events with energies above 300 MeV from within 1.0\degr\ of the pulsar over the date range 2008 August 4 to 2018 February 14, and applied a zenith angle cut of 100\degr. We fitted the timing model parameters using an unbinned Markov Chain Monte Carlo (MCMC) Maximum Likelihood technique (see \citealt{2PC} and \citealt{pc15}); \texttt{PINT} \citep{PINT} 
includes an open-source implementation of this technique, called \texttt{event\_optimize}. The underlying MCMC engine is {\tt emcee} \citep{emcee},  which uses affine transforms to efficiently explore high-dimensional parameter spaces 
and map out parameter confidence regions, even when they are highly correlated.

For the pulse template we used a three-Gaussian model and for the
necessary photon weights ($w_j$), we used an empirical calculation based on a typical $\gamma$-ray pulsar spectrum and an approximation to the LAT point spread function as a function of energy and angular offset from the pulsar position \citep{bruel19}. The weights represent the probability that the photon originated from the pulsar, as opposed to other point sources or the diffuse background.

\begin{figure}
\plottwo{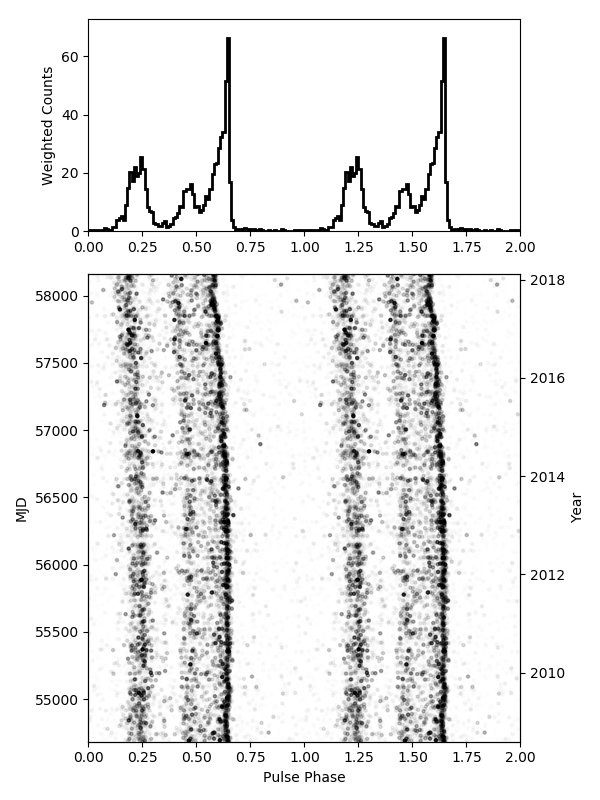}{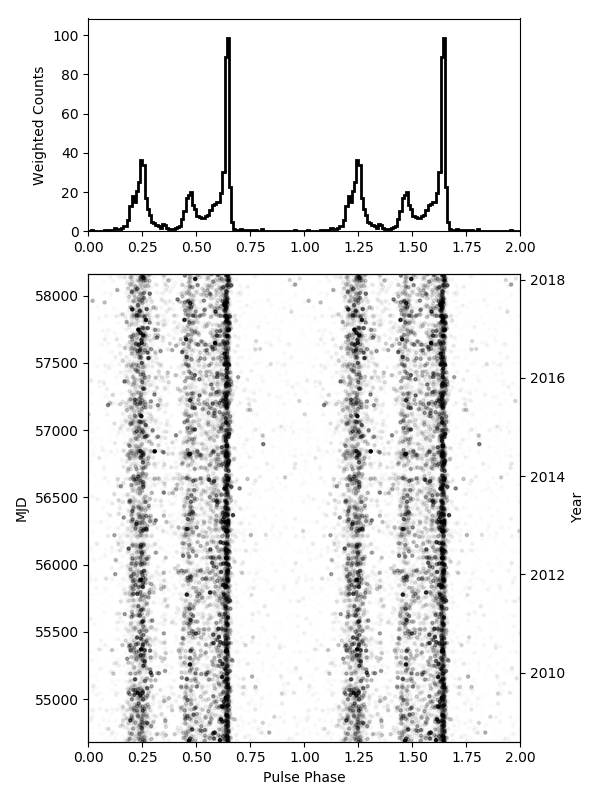}
\caption{\textit{Fermi}-LAT phaseograms using the unmodified 2PC (left) and final MCMC (right) timing models. Photons above 300 MeV from within 1\degr\ of the pulsar direction are plotted with a grayscale based on the probability that the photon came from the source.\label{fig:phaseogram}}
\end{figure}

In this calculation, we fit for the spin frequency and its derivative, 
the position and proper motion of the pulsar, and the binary parameters using
the \texttt{ELL1} model appropriate for nearly circular orbits \citep{ELL1}. The 
maximum likelihood model along with uncertainties estimated from the posterior probability
distribution are shown in Table \ref{tab:timing}. A phaseogram showing the full 
\textit{Fermi} \replaced{mission}{dataset} before and after updating the timing model is shown in Figure 
\ref{fig:phaseogram}.

\begin{figure}
\plotone{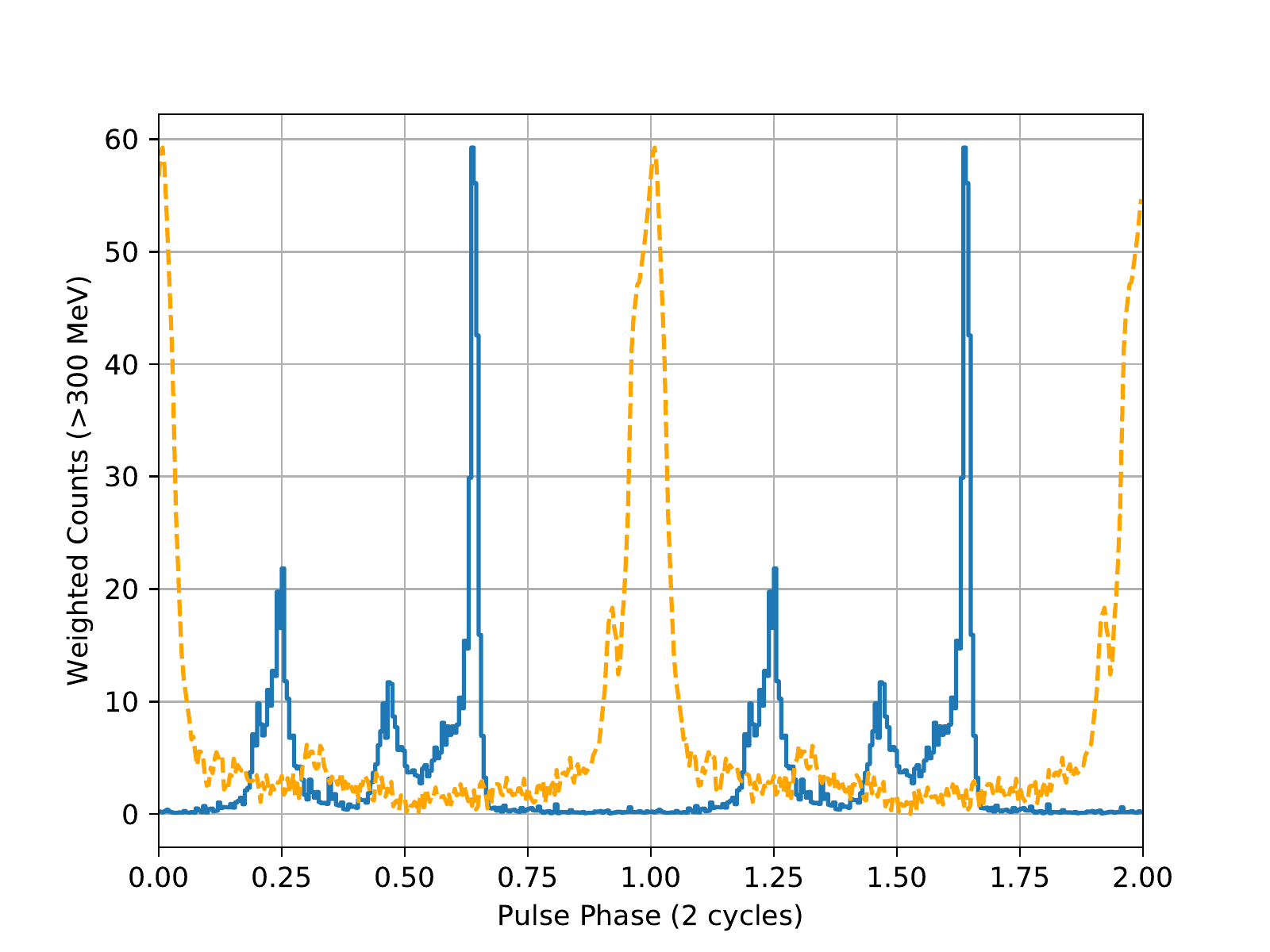}
\caption{Phase-aligned pulse profiles from {\it Fermi} LAT ($>300$ MeV, blue) and {\it Nan\c{c}ay Radio Telescope} (1408 MHz, dashed orange line, arbitrary amplitude). The phase and separation of the {\it Fermi} LAT peaks are consistent with what was reported in 2PC.\label{fig:fermi}}
\end{figure}

Throughout this work, we adopt the same phase 0 definition as was used in 2PC, specified
by the TZR parameters (which define a \replaced{TOA}{pulse time of arrival} that has phase 0.0 according to the model) in Table \ref{tab:timing}. A plot of the \textit{Fermi} pulse profile, 
phase-aligned with the 1408 MHz radio profile is shown in Figure \ref{fig:fermi}.

\begin{deluxetable}{lr}
\caption{Maximum Likelihood {\it Fermi} LAT Timing Model\label{tab:timing}}
\tablehead{\colhead{Parameter} & \colhead{Value}}
\startdata
Right Ascension ($\alpha$, J2000)                 & 12$^{\rm h}$31$^{\rm m}$11\fs3131(1) \\
Declination ($\delta$, J2000)                     & $-$14\arcdeg11\arcmin43\farcs642(3) \\
Proper Motion in R.A. ($\mu_\alpha \cos \delta$, mas yr$^{-1}$)  & $-$61.5(7) \\
Proper Motion in Decl. ($\mu_\delta$, mas yr$^{-1}$)             &  6.6(3) \\
Epoch of position (MJD)                           &  55000.0 \\
Pulse frequency ($\nu$, Hz)                           & 271.453019624388(4) \\
Frequency derivative ($\dot{\nu}$, s$^{-2}$)                & $-1.66705(4) \times 10^{-15}$ \\
Epoch of frequency                                &  55000.0 \\
Dispersion Measure (cm$^{-3}$)                    & 8.09 \\
Binary Model                                      & ELL1 \\
Binary Period ($P_B$, d)                             &   1.8601438845(2)\\
Semimajor axis ($a_1 \sin(i)$, lt\,s)                    &      2.042625(1)\\
Epoch of ascending node ($T_\mathrm{asc}$, MJD)        & 55015.1534653(2)\\
First Laplace parameter ($10^{-7}\,e \sin \omega$)         &    $-$10(7) \\
Second Laplace parameter ($10^{-7}\,e \cos \omega$)        &   $-$3(8) \\
Timescale & TDB \\
Solar System Ephemeris & DE421 \\
TZRMJD    &     55242.107268755294338103444 \\
TZRFRQ    &     1408.0 \\
TZRSITE   &     \texttt{ncy} \\
\enddata
\end{deluxetable}
 
\subsection{\nicer{} Pulsation Search}

Using the timing model from Table \ref{tab:timing}, we searched for pulsations in the \nicer{} data.  We note that a 1.0 second offset is present in the raw \nicer\ science data, due to a time assignment error in the on-board software. This was discovered early in the mission by comparison of absolute arrival times of X-ray and radio pulsations from the Crab Pulsar, PSR B1937+21 and PSR B1821$-$24, in support of the SEXTANT pulsar navigation experiment \citep{SEXTANTGNC}. In the current pipeline processing, this correction is applied by setting the FITS header parameter \texttt{TIMEZERO} to $-1.0$\,s in the data distributed by the HEASARC.

Pulse phases for each photon were computed using the \texttt{photonphase} code in \texttt{PINT} 
For each photon, the position of \nicer\ is interpolated from the orbit file (which has state vector points at 10 second intervals) and used to compute the Solar System time delays in the computation of the model phase.

\begin{figure}
\plotone{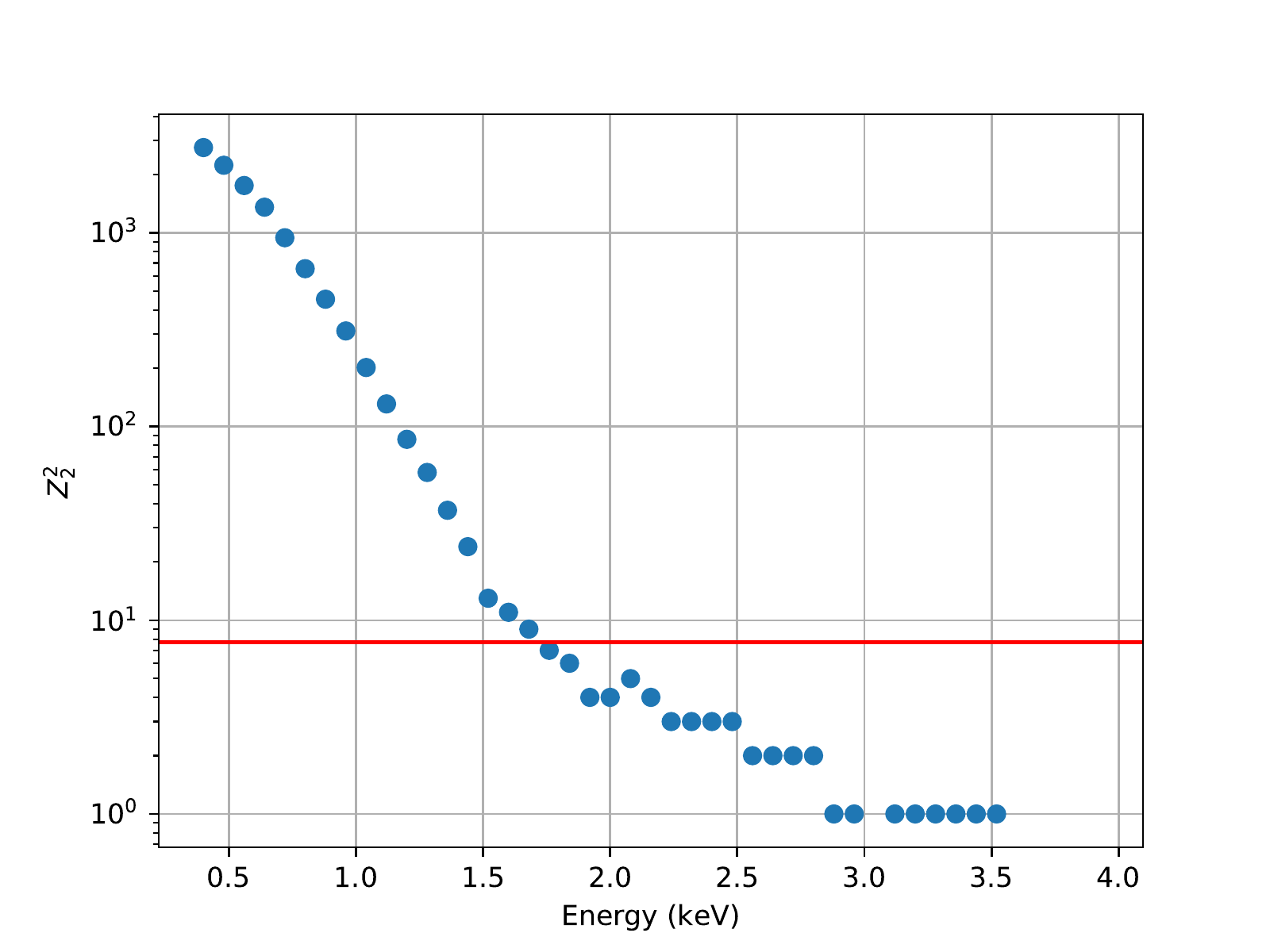}
\caption{\label{fig:z2}$Z^2_2$ test from \nicer{} data as a function of low energy cut. The horizontal red line indicates the threshold for a 90\% confidence detection of pulsations.}
\end{figure}

With pulse phases assigned, we computed the H-test \citep{Htest} and detected pulsations with a 
significance of 55.3$\sigma$ ($H=3193$) for our initial energy cuts of 0.3--8 keV.  The maximal H-test ($H = 5007$, corresponding to 69.7$\sigma$) is obtained when selecting an energy range of 0.31 to 1.51 keV, suggesting that the pulses are thermal, as expected from the soft spectrum
and broad pulses.  To look for evidence of a hard pulsed component, we computed the $Z^2_2$ test (appropriate for the smooth pulse profile observed; \citealt{Htest}) for different minimum energy cuts, as shown in Figure \ref{fig:z2}.  We see no evidence for a  pulsed signal above 1.7 keV.

\begin{figure}
\plotone{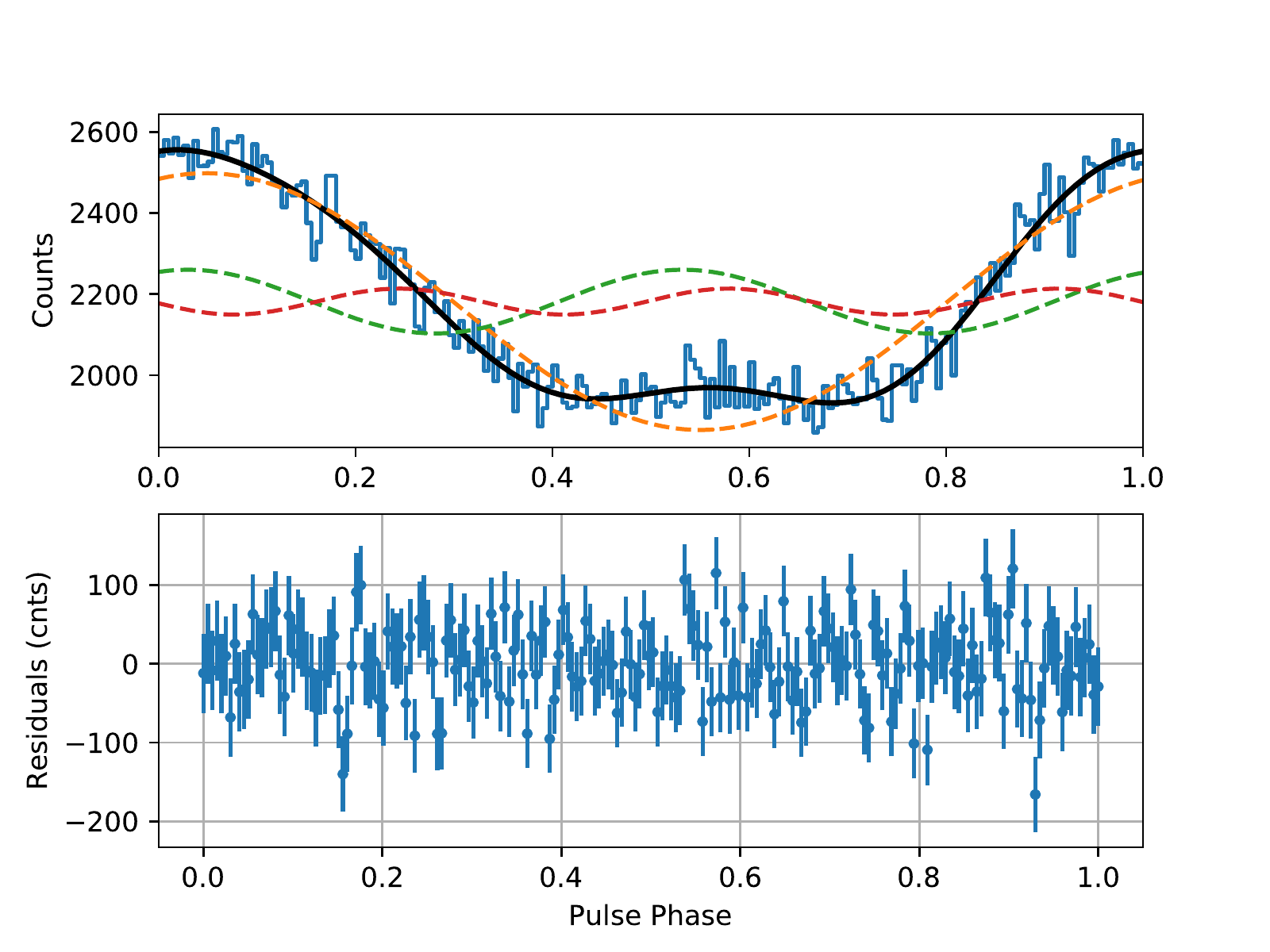}
\caption{\label{fig:harms} \nicer{} pulse profile (0.35--1.5 keV; \added{200 phase bins}) decomposed into three harmonically-related sinusoids. The solid black curve is the sum of the three dashed components. The bottom panel shows the residuals to the full fit.}
\end{figure}

\begin{figure}
\plotone{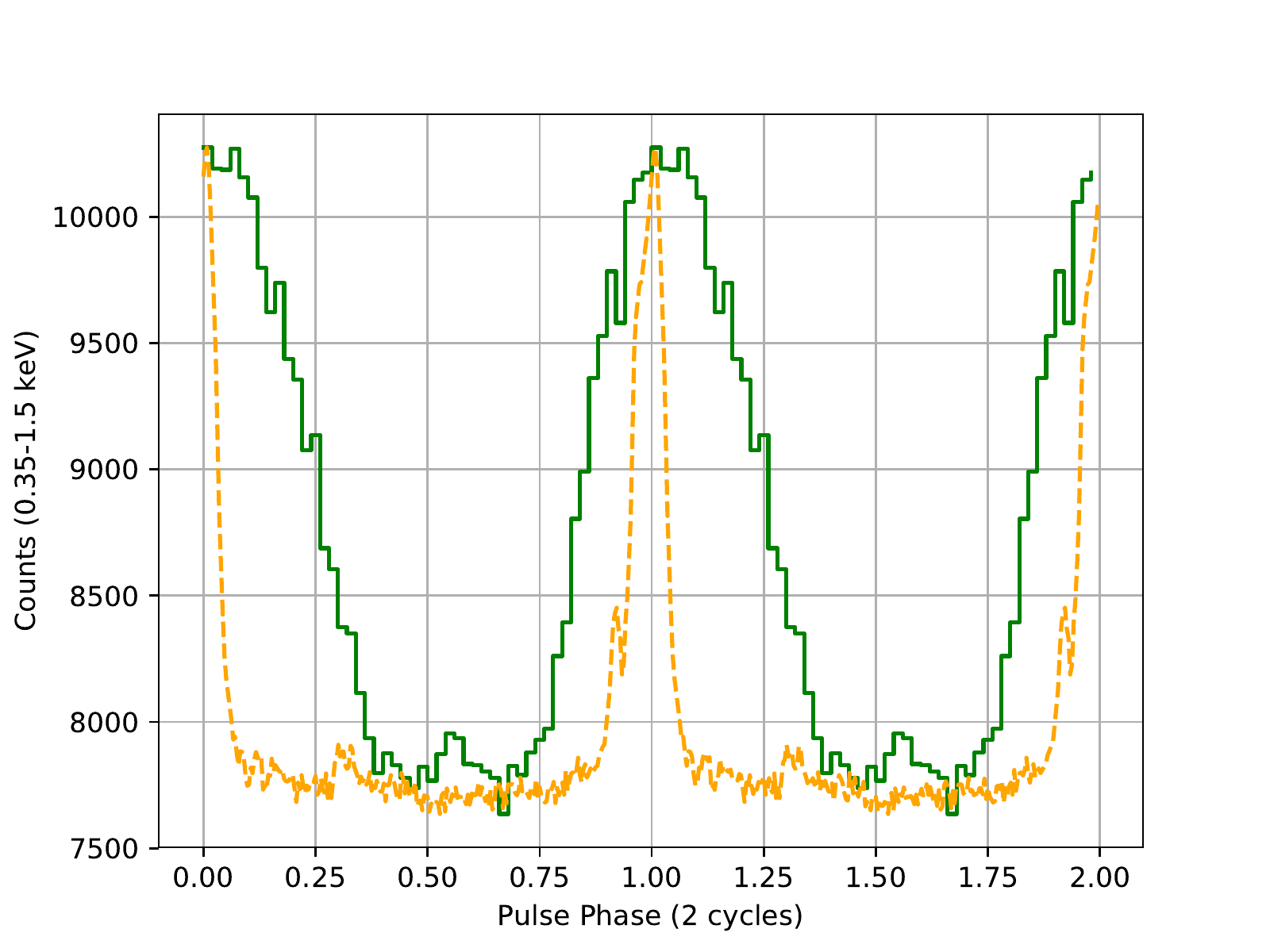}
\caption{Phase-aligned pulse profiles from \nicer{} (0.35--1.5 keV, green; \added{50 bins per phase}) and the {\it Nan\c{c}ay Radio Telescope} (1408 MHz, dashed orange line, arbitrary amplitude). \label{fig:nicer}}
\end{figure}

We computed the amplitudes of the first 20 Fourier components from the unbinned pulse phases and found that only the fundamental and first two harmonics are significant above the $2\sigma$ level. The pulse profile with the three harmonic decomposition is shown in Figure \ref{fig:harms}, while the X-ray and radio phase-aligned profiles 
are shown in Figure \ref{fig:nicer}. The pulsed count rate is 0.055 c s$^{-1}$ (0.35--1.5 keV), while the unpulsed background is 0.422 c s$^{-1}$. The unpulsed background contains contributions from radiation background, diffuse X-ray background, detector \replaced{backgrounds}{noise}, and unpulsed emission from the source. The spectral analysis presented in \S\ref{sec:spec} gives a total source count rate of 0.145 c s$^{-1}$  (0.35--1.5 keV), yielding a pulsed fraction of 38\% in that 
energy band. There is a substantial systematic uncertainty in the pulsed fraction, which is dominated by the uncertainty in the background model (since the background gives 2/3 of the total count rate observed). 
Archived {\it XMM-Newton} imaging observations ($\sim 30$ ks, \citealt{rrc+11}) show that there are no strong contaminating sources in the \nicer\ field of view, so the uncertainty comes from cosmic variance in the diffuse X-ray background and inaccuracies in the model of the radiation background.

\section{Spectral analysis}
\label{sec:spec}
The event filtering used to generate the spectrum of \psr{} differs from that described in \S\ref{sec:obs}. Indeed, a more stringent filtering permits minimizing non-astrophysical background (particle flaring, optical loading on the detectors, etc.), especially for faint sources such as \psr.  In addition to the filtering described in \S\ref{sec:obs}, we exclude observations where the Sun angle is $<80\degr$. 

In addition, we apply a filter based on the cutoff rigidity (\texttt{COR\_SAX}) and a housekeeping parameter ({\tt FPM\_OVERONLY\_COUNT}) that counts detector overshoots (large energy depositions in the detector), which are strongly correlated with the radiation background level.
Specifically, we filter out time intervals with ${\tt FPM\_OVERONLY\_COUNT}>1.0$ or ${\tt FPM\_OVERONLY\_COUNT}>\left(1.52\times{\tt COR\_SAX}^{-0.633}\right)$. This empirical relation maximizes exposure, even at low cut-off rigidities (${\tt COR\_SAX}$), as long as the ${\tt FPM\_OVERONLY\_COUNT}$ is not too large.  This filtering results in 723.7 ks of exposure.

\begin{deluxetable}{llccc}
\tablecaption{Results of the \nicer{} spectral analysis for \psr{} with simple models\label{tab:spec}}
\tablehead{
\colhead{Component} & \colhead{Parameter} & \colhead{BB+PL} & \colhead{BB+BB} & \colhead{\tt nsatmos} 
}
\startdata
{\tt tbabs}    & $N_{\rm H}$ ($\ee{20}\percmsq$) & 5\ud{7}{3} & 6\ud{3}{6}  & 0.8\ppm0.4\\
{\tt Gaussian} & $E_{\rm G}$ (keV)      & 0.577\ppm0.004 & 0.575\ppm0.004 & 0.576\ppm0.004\\
               & $\sigma_{\rm G}$ (keV) & 0.030\ud{0.008}{0.009} & 0.036\ud{0.008}{0.009} & 0.025\ud{0.008}{0.009}\\
               & Norm ($\ee{-5}$ ph $\percmsq\persec$)      & 2.8\ud{3.5}{0.9} & 3.8\ud{2.2}{2.0} & 1.6\ppm0.2 \\
{\tt bbodyrad} & $kT$ (eV)         & 136\ud{7}{14} & 44\ud{32}{8} & -- \\
               & Norm ($R^{2}_{\rm km}/D^{2}_{\rm 10\,kpc}$)   & 35\ud{60}{11} & 43500\ud{556000}{43200} & --\\
{\tt bbodyrad} & $kT$ (eV)         & -- & 133\ud{16}{6} & -- \\
               & Norm ($R^{2}_{\rm km}/D^{2}_{\rm 10\,kpc}$)   & -- & 59\ud{41}{37} & -- \\
{\tt powerlaw} & $\Gamma_{\rm PL}$ & 5.0\ud{3.9}{1.1} & -- & --\\
               & Norm (ph $\,{\rm keV}^{-1}\percmsq\persec$)  & (8.1\ud{6.5}{7.3})\tee{-6} & -- & --\\
{\tt nsatmos}\tablenotemark{(1)}  & $kT$ (eV)        & -- & -- & 51\ppm2\\
               & Norm              & -- & -- & 0.10\ppm0.02\\
\hline
\multicolumn{2}{c}{$F_{\rm 0.3-2.0\,keV}$ ($\ee{-13}\cgsflux$) } & 1.61\ud{0.01}{0.59} & 1.61\ud{0.02}{1.60} & 1.62\ud{0.01}{0.05} \\
\multicolumn{2}{c}{\Chisqred\,(d.o.f.)} & 1.63 (71) & 1.52 (71) & 1.54 (73)
\enddata
\tablecomments{BB={\tt bbodyrad}, PL={\tt powerlaw}. All errors reported are at 90\% confidence.}
\tablenotetext{1}{For the {\tt nsatmos} model, we fixed the parameters $R=11\km$, $M=1.4\msun$ and $d=420$~pc. The {\tt nsatmos} normalization corresponds to the fraction of neutron star emitting.}
\end{deluxetable}

The background spectrum was generated from a grid of \nicer{} blank-sky spectra corresponding to the blank-sky pointings of \textit{Rossi X-ray Timing Explorer} \citep{Jahoda06}. This grid of spectra is populated with observed spectra in various space-weather observing conditions (Gendreau et al. in prep.). The background spectrum is generated by combining these blank-sky spectra weighted according to space-weather conditions and magnetic cutoff rigidities common to both the pulsar and background-fields observations.

We used the spectrum in the 0.3--1.5\,keV energy range (the optimal range from the timing analysis), which resulted in about 122,000 source counts (out of 356,000 total counts).  Above 1.5 keV, the background count rate in each spectral bin dominates the source spectrum count rate by over two orders of magnitude. Below 0.3 keV, the characterization of the noise peak (due to optical loading) remains uncertain, even at Sun angles $>80\degr$.  We add 2\% systematic in each spectral bin to account for uncertainties in the calibration as estimated from observed residuals in fits to the \nicer{} Crab spectrum. Finally, we used the \nicer\ response files version 0.06, but the ancillary response file was re-scaled by a factor 49/52, to account for the three detectors excluded.

\begin{figure}[t!]
\centering
\includegraphics[width=0.9\textwidth]{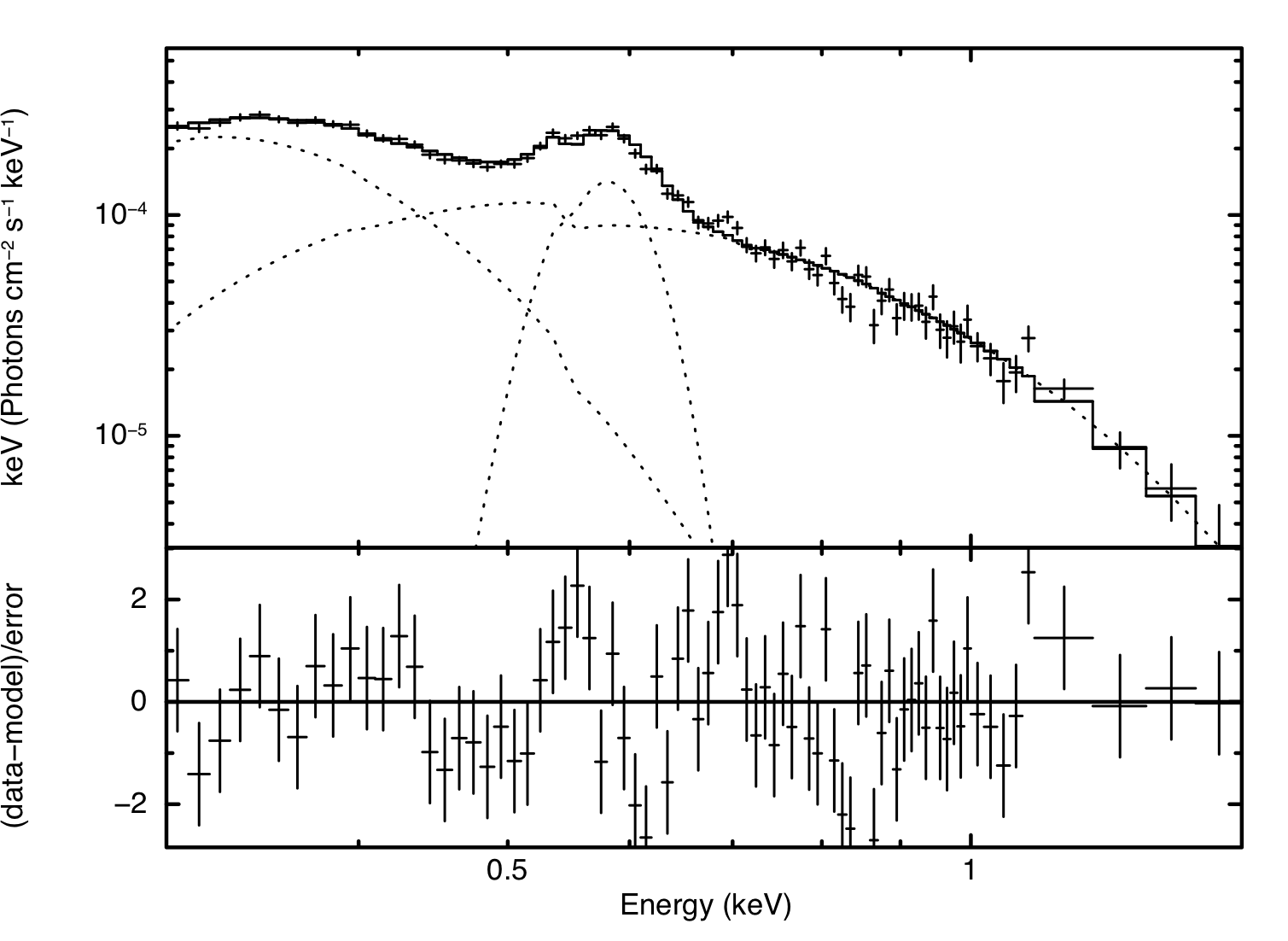} 
\caption{\nicer\ unfolded spectrum of PSR J1231--1411, with the double-blackbody model and including a Gaussian emission line at 0.575 keV. The dotted lines indicate the individual components (cold blackbody dominating at low energies, and the hot blackbody dominating at high energies). The bottom panel shows the residuals.}
\label{fig:spec1}
\end{figure}

To model the Galactic absorption, we used the {\tt tbabs} model, with the {\tt VERN} cross-sections \citep{verner96} and {\tt WILM} abundances \citep{wilms00}. The spectral continuum is modeled with either a power law (model {\tt powerlaw}) or a black body (model {\tt bbodyrad}), or a combination of these, as described below. We also tried a non-magnetic neutron star atmosphere model {\tt nsatmos}, as often employed to describe the spectra of MSPs \citep[e.g., for PSR~J0437--4715, ][]{2013ApJ...762...96B,guillot16}, since they are expected to have magnetic fields of the order of $10^{8}$--$10^{9}$~G.  Finally, we add a Gaussian line at $E\approx0.57$\,keV (all parameters are fitted) to account for a foreground feature, unrelated to the source, and thought to be \ion{O}{7} 
\replaced{feature}{emission} caused by Solar wind charge exchange or originating in the local hot bubble \citep[e.g.,][]{gupta09,galeazzi14}.

First, using a simple absorbed {\tt powerlaw}, with or without a {\tt Gaussian} results in unacceptable fits (\Chisqred$\sim$3 and 17, respectively). Moreover, the  best-fit photon index, $\Gamma\sim 4$ is reminiscent of blackbody-like components.  Adding the {\tt Gaussian} feature to a  single {\tt bbodyrad} model improves the \replaced{fits}{goodness-of-fit} statistics from \Chisqred$\sim6.4$ to $\sim2.8$. However, the fit quality \replaced{is still}{remains} poor, and the structures observed in the residuals warrant the addition of a second spectral component.  

\begin{figure}[t!]
\centering
\includegraphics[width=0.9\textwidth]{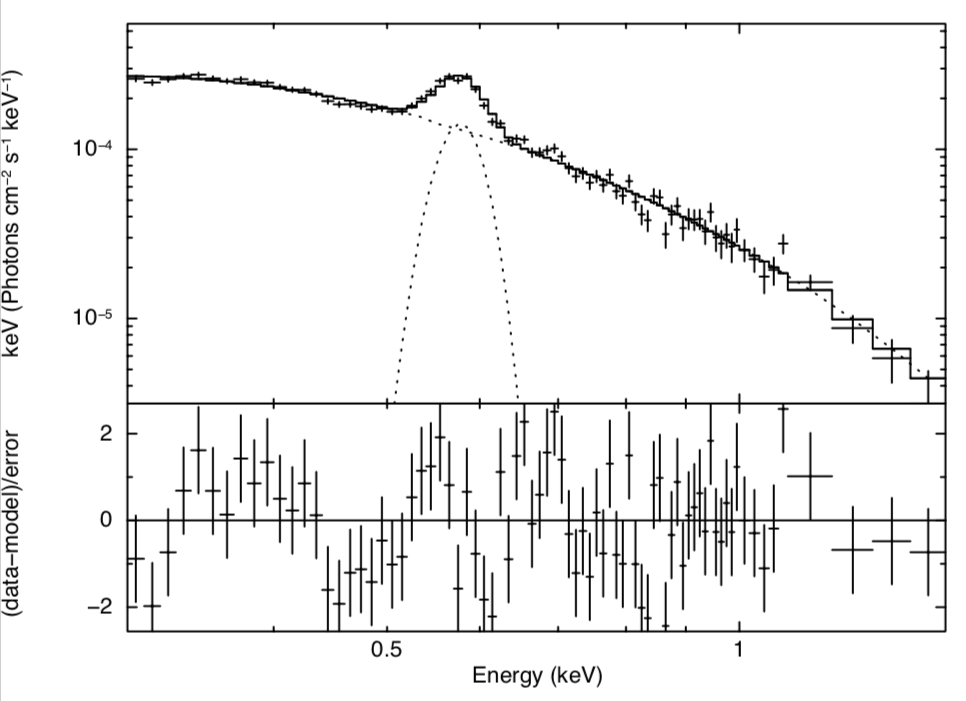} 
\caption{\nicer\ unfolded spectrum of PSR J1231$-$1411, fitted with the {\tt nsatmos} model, where a {\tt Gaussian} component has also been added, as in Figure~\ref{fig:spec1}. The bottom panel shows the residuals.}
\label{fig:spec2}
\end{figure}

In Table~\ref{tab:spec}, we therefore report the spectral fits of \psr{} with a double-blackbody model (see Figure \ref{fig:spec1}) and with a {\tt bbodyrad+powerlaw} model (together with the {\tt Gaussian} as in the model above). In the latter, the photon index of the power law is also extremely soft ($\Gamma\sim 5$), which favors the double-blackbody model. In these two models, the normalizations are poorly constrained, especially for the cold {\tt bbodyrad} component. Finally, we also report the fit with the {\tt nsatmos} (fixing $M=1.4$\,$M_{\odot}$, $R=11$\,km and $d=420$\,pc), which does not require the addition of a second continuum component (see Figure \ref{fig:spec2}). Note that any uncertainties in the distance would be directly incorporated into the uncertainties of the normalization. The fit is insensitive to freeing the radius and mass. Finally, because of the dominating background above 1.5\,keV, we cannot determine the presence of a hard X-ray tail, as observed for other MSPs (e.g., PSR~J0437--4715, \citealt{zavlin02, guillot16}).

\section{Summary and discussions}

The detection of thermal pulsations from \psr{} was an important pre-launch goal for \nicer{}, which has now been realized. This has spurred a large investment of observing time that will continue to build up a high
signal-to-noise energy-resolved pulse profile to enable lightcurve modeling to constrain the mass and radius of the neutron star in this system \citep{2016EPJA...52...37B}.  We found a predominantly thermal spectrum for \psr{}, and an X-ray luminosity,  $L_{\rm X} = \left(3.42\ud{0.02}{0.11}\right)\times10^{30}\,{\rm erg s^{-1}}$ (assuming $d=420$\,pc), \added{i.e., $\sim7\times10^{-4}\,\dot{E}$}, typical of thermally emitting MSPs \citep{forestell14,lee18}, and consistent with the X-ray luminosity reported previously \citep{rrc+11}.

\subsection{Emission Geometry}

The $\gamma$-ray and radio pulse profiles of this pulsar have been studied previously,
providing information about the geometry of the emitting regions \citep{Johnson2014}. 
In these models, the radio emission is assumed to be from a hollow cone centered on the magnetic polar cap, while multiple models for the
location of the $\gamma$-ray emitting region are tested: outer gap (OG), two-pole caustic (TPC), and pair-starved polar cap (PSPC). The model light curves are computed over two parameters, $\alpha$ (the angle between the spin axis and the magnetic axis), and $\zeta$ (the angle between the spin axis and the line of sight).  Since the $\gamma$-ray model light curve phases are relative to the (unknown) magnetic pole phase and the \textit{Fermi} light curve phase is determined relative to that of the radio peak, an additional free parameter for the phase offset between the magnetic pole and the radio peak phase is derived from the fits of the models to the data.  The derived phase offsets, $\Delta\phi$, for most of the pulsars are positive, indicating that the radio emission comes from an altitude above the neutron star surface. {\citet{Johnson2014} provide the best-fit location of the magnetic pole, without uncertainty, as $\Phi_{\mu}$, and in the case of PSR J1231$-$1411 they used the radio profile from \citet{2PC} which put the radio peak at phase 0 allowing us to equate $\Phi_{\mu}$ and $\Delta\phi$ for this pulsar.}

In the case of \psr, the \nicer{} X-ray light curve (see Figures \ref{fig:harms} and \ref{fig:nicer}) can be used to measure $\Delta\phi$, assuming
that the X-rays come from a region on the surface centered on the magnetic pole.
The peak of the fundamental harmonic in Figure \ref{fig:harms} is at a phase of  $\Delta\phi_\mathrm{obs} = 0.05$. This can provide constraints on both the radio emission height and on the $\gamma$-ray models.

Radio pulses occurring at a radius $r$ will arrive at an observer at a phase ahead of that of the magnetic pole due to aberration and retardation \citep{Dyks2004}, each of which produce a phase shift $\Delta\phi = -r/R_{\rm LC}$ where $R_{\rm LC} = c/\Omega$. However, this is balanced by the backward shift of the polar cap caused by the rotational sweep back of field lines near the light cylinder of $\Delta\phi \sim 0.2 (r/R_{\rm LC})^{1/2}$ \citep{DyksHarding2004}.  The total phase shift is $\Delta\phi_{\rm tot} = -2r/R_{\rm LC} + 0.2 (r/R_{\rm LC})^{1/2}$ .  Equating this with the measured phase shift $\Delta\phi_\mathrm{obs}$ between the main X-ray and radio peaks indicates that the radio emission radius is $r \sim 0.047\,R_{\rm LC} = 8.3$ km, essentially at the neutron star surface.  

The measured phase shift $\Delta\phi_{\rm obs}$ can also constrain the $\gamma$-ray model if it is associated with the $\Delta\phi$ from the fits.  From the fits of \citet{Johnson2014} for \psr, the OG model gives $\alpha = 88^\circ$, $\zeta = 67^\circ$ and $\Delta\phi = 0.056$, while the TPC model gives $\alpha = 26^\circ$, $\zeta = 69^\circ$ and $\Delta\phi = 0.022$. The PSPC model provides a poor fit to \psr{} given that its very sharp gamma-ray peaks that lag the main radio peak are in conflict with the predictions of that model, so we don't consider it further. In \citet{Johnson2014}, the OG model for the combined $\gamma$-ray-radio fit also has a slightly higher likelihood than the TPC model because of the better match to the $\gamma$-ray peaks. The $\Delta\phi$ from the X-ray is consistent with this preference. 

\citet{Bez2018} have also performed fits of the same models to the $\gamma$-ray light curve alone, obtaining $\alpha = 82^\circ$, $\zeta = 65^\circ$ and $\Delta\phi = 0.040^{+0.016}_{-0.008}$ for the OG model and $\alpha = 71^\circ$, $\zeta = 59^\circ$ and $\Delta\phi = 0.072^{+0.016}_{-0.008}$ for the TPC model.  So again, the \nicer-measured phase 
is more consistent with the OG model.

However, in both \citet{Johnson2014} and \citet{Bez2018}, the
OG fits have the pulsar being a nearly orthogonal rotator ($\alpha$ near 90$^\circ$). This geometry
could have difficulty matching the radio profile because it tends to predict 
a radio interpulse, which is not observed. If $(180-\alpha) - \theta_{\rm R} > \zeta$, where $\theta_{\rm R}$ is the angular size of the radio emission cone, then the observer will miss the second radio peak and see only one radio pulse.  With the constraint on the radio emission height above, the angular size of the radio emission cone of \psr{} at $r = 0.047\,R_{\rm LC}$ is $\theta_{\rm R} \sim 1.5(r/R_{\rm LC})^{1/2} = 18^\circ$.  This estimate indicates that the OG fits of both \citet{Johnson2014} and \citet{Bez2018} will predict only one visible radio pulse, consistent with what is observed. The two differ in that \citet{Bez2018} fits only the $\gamma$-ray light curve.

For a nearly orthogonal rotator, if the polar caps are nearly antipodal (as expected from a dipole geometry) and of similar size and temperature, there should be X-ray peaks of similar 
magnitude separated by about 180$^\circ$, which also is not observed. The X-ray profile 
appears to show emission from both polar caps, but indicates a moderate $\alpha$ and large $\beta = \alpha - \zeta$ to reproduce the large amplitude ratio between the peaks. Of course, the geometry could
be more complicated than antipodal hot spots with similar properties. The detailed
pulse profile modeling that is ongoing as part of the effort to constrain the neutron
star radii of thermally-emitting MSPs will provide insight into the hot spot geometry \citep{2016EPJA...52...37B}.

So, while the phase offset between the radio and X-ray peaks supports the OG models for the
$\gamma$-ray emission, the X-ray pulse profile shape (primarily the large asymmetry of the peaks)
tends to support a smaller inclination angle. Additional modeling of the $\gamma$-ray and radio light curves using the X-ray determined phase shift, $\Delta\phi_\mathrm{obs}$ as a prior, may
resolve this discrepancy.

\subsection{Spectral analysis}

Although the spectral analysis is somewhat limited by the faintness of \psr{} (only $\sim 1/3$ of observed counts), its spectrum has a predominantly thermal origin. When using a blackbody, a second component (blackbody or very steep $\Gamma\sim 5$ power law) is required. However, when using a neutron star atmosphere model ({\tt nsatmos}), which has a somewhat harder tail than a Planck function, no additional continuum component is required to obtain an equally good fit compared to the two-component models (see Table~\ref{tab:spec}).

A neutron star atmosphere is most likely to best describe the thermal emission from the polar caps of a MSP such as \psr{}. However, in that interpretation, the polar cap of this pulsar would have a temperature ($51\ppm2$\,eV) lower than typically observed for other MSPs ($\gtrsim 90$ up to $\sim$150 eV, see \citealt{2009ApJ...703.1557B,2013ApJ...762...96B}). In addition, polar caps covering $\sim 10\%$ of the total surface would be larger than expected for a neutron star with a $\sim$3-ms spin period (for which the polar cap radius would be $\sim$2--3~km, \citealt{arons81,dermer94}). Attempts to fit the data with a double-{\tt nsatmos} model resulted in an unstable fit as our data set does not require the addition of a component to the single {\tt nsatmos} model.

On the other hand, the double-blackbody model could in principle be an adequate description of the data, where each component represents the emission from the two polar caps, a hot polar cap with a colder annulus around it, or a hot polar cap and the remainder of the colder surface.  The latter is supported by the effective areas of these two components: $\sim 8$\,km for the 44\,eV blackbody and $\sim320$\,m for the 133\,eV blackbody (see Table~\ref{tab:spec}), but we keep in mind that the normalizations of these two components are poorly constrained.  For PSR~J0437--4715, the brightest and nearest MSP ($\times 3$ closer than \psr), the cold emission from the entire surface of that MSP has a temperature of $\sim$ 30--35\,eV (e.g., \citealt{2013ApJ...762...96B,guillot16}).  A more precise determination of the temperatures of the polar caps will likely arise from the full pulse profile modeling analysis, and will be presented in an upcoming publication.

Using {\tt nsatmos}, we find a redshifted temperature of $40\pm2$~eV (assuming a 1.4-\msun, 11-km neutron star), which is inconsistent with the value obtained from \textit{XMM} data ($61^{+6}_{-12}$~eV, \citealt{rrc+11}). This might be caused by the fact that an additional power-law component is required by the \textit{XMM} data. Our \nicer{} spectrum, however, has a dominating background above 1.5\, keV which prevents significant constraints on this component.
The overall flux reported previously is consistent with the one listed in Table~\ref{tab:spec}.

\acknowledgments

This work was supported by NASA through the \nicer{} mission and the 
Astrophysics Explorers Program. This research has made use of data and/or 
software provided by the High Energy Astrophysics Science Archive Research 
Center (HEASARC), which is a service of the Astrophysics Science Division at 
NASA/GSFC and the High Energy Astrophysics Division of the Smithsonian 
Astrophysical Observatory.  We thank Faith Huynh for her excellent \nicer{} 
quicklook plotting software, and Tyrel Johnson for useful discussions. 
\nicer{} work at NRL is supported by NASA. SG acknowledges the support of the
CNES. This research has made use of the NASA Astrophysics Data System (ADS) and
the arXiv.

The \textit{Fermi} LAT Collaboration acknowledges generous ongoing support
from a number of agencies and institutes that have supported both the
development and the operation of the LAT as well as scientific data analysis.
These include the NASA and the Department of Energy in the United States, the Commissariat \`a l'Energie Atomique
and the Centre National de la Recherche Scientifique / Institut National de Physique
Nucl\'eaire et de Physique des Particules in France, the Agenzia Spaziale Italiana
and the Istituto Nazionale di Fisica Nucleare in Italy, the Ministry of Education,
Culture, Sports, Science and Technology (MEXT), High Energy Accelerator Research
Organization (KEK) and Japan Aerospace Exploration Agency (JAXA) in Japan, and
the K.~A.~Wallenberg Foundation, the Swedish Research Council and the
Swedish National Space Board in Sweden.
 
Additional support for science analysis during the operations phase is gratefully
acknowledged from the Istituto Nazionale di Astrofisica in Italy and the Centre
National d'\'Etudes Spatiales in France. This work performed in part under DOE
Contract DE-AC02-76SF00515.

The National Radio Astronomy Observatory is a facility of the National Science Foundation operated under cooperative agreement by Associated Universities, Inc.  SMR is a CIFAR Fellow and is supported by the NSF Physics Frontiers Center award 1430284.

%

\vspace{5mm}
\facilities{\nicer, \textit{Fermi}}


\software{
\texttt{astropy} \citep[ascl:1304.002]{2013A&A...558A..33A}, \texttt{PINT} \replaced{(ascl:1902.007, \url{https://github.com/nanograv/pint})}{\citep[ascl:1902.007]{PINT}},
HEAsoft (\url{https://heasarc.nasa.gov/lheasoft/}, ascl:1408.004), 
\texttt{emcee} \replaced{(ascl:1303.002, \url{https://github.com/dfm/emcee})}{\citep[ascl:1303.002]{emcee}}
}

\bibliographystyle{aasjournal}
\bibliography{nicer}

\begin{thebibliography}{}
\expandafter\ifx\csname natexlab\endcsname\relax\def\natexlab#1{#1}\fi
\providecommand{\url}[1]{\href{#1}{#1}}
\providecommand{\dodoi}[1]{doi:~\href{http://doi.org/#1}{\nolinkurl{#1}}}
\providecommand{\doeprint}[1]{\href{http://ascl.net/#1}{\nolinkurl{http://ascl.net/#1}}}
\providecommand{\doarXiv}[1]{\href{https://arxiv.org/abs/#1}{\nolinkurl{https://arxiv.org/abs/#1}}}

\bibitem[{{Abdo} {et~al.}(2013){Abdo}, {Ajello}, {Allafort}, {Baldini},
  {Ballet}, {Barbiellini}, {Baring}, {Bastieri}, {Belfiore}, {Bellazzini}, \&
  et~al.}]{2PC}
{Abdo}, A.~A., {Ajello}, M., {Allafort}, A., {et~al.} 2013, \apjs, 208, 17,
  \dodoi{10.1088/0067-0049/208/2/17}

\bibitem[{{Alpar} {et~al.}(1982){Alpar}, {Cheng}, {Ruderman}, \&
  {Shaham}}]{1982Natur.300..728A}
{Alpar}, M.~A., {Cheng}, A.~F., {Ruderman}, M.~A., \& {Shaham}, J. 1982, \nat,
  300, 728, \dodoi{10.1038/300728a0}

\bibitem[{{Arons}(1981)}]{arons81}
{Arons}, J. 1981, \apj, 248, 1099, \dodoi{10.1086/159239}

\bibitem[{{Astropy Collaboration} {et~al.}(2013){Astropy Collaboration},
  {Robitaille}, {Tollerud}, {Greenfield}, {Droettboom}, {Bray}, {Aldcroft},
  {Davis}, {Ginsburg}, {Price-Whelan}, {Kerzendorf}, {Conley}, {Crighton},
  {Barbary}, {Muna}, {Ferguson}, {Grollier}, {Parikh}, {Nair}, {Unther},
  {Deil}, {Woillez}, {Conseil}, {Kramer}, {Turner}, {Singer}, {Fox}, {Weaver},
  {Zabalza}, {Edwards}, {Azalee Bostroem}, {Burke}, {Casey}, {Crawford},
  {Dencheva}, {Ely}, {Jenness}, {Labrie}, {Lim}, {Pierfederici}, {Pontzen},
  {Ptak}, {Refsdal}, {Servillat}, \& {Streicher}}]{2013A&A...558A..33A}
{Astropy Collaboration}, {Robitaille}, T.~P., {Tollerud}, E.~J., {et~al.} 2013,
  \aap, 558, A33, \dodoi{10.1051/0004-6361/201322068}

\bibitem[{Atwood {et~al.}(2013)Atwood, Albert, Baldini, {et~al.}}]{P8}
Atwood, W.~B., Albert, A., Baldini, L., {et~al.} 2013, in Proc. of the 4th
  International Fermi Symposium, eConf C121028, (arXiv:1303.3514)

\bibitem[{{Bezuidenhout} {et~al.}(2018){Bezuidenhout}, {Venter}, {Seyffert}, \&
  {Harding}}]{Bez2018}
{Bezuidenhout}, M.~C., {Venter}, C., {Seyffert}, A.~S., \& {Harding}, A.~K.
  2018, arXiv e-prints.
\newblock \doarXiv{1808.09762}

\bibitem[{{Bogdanov}(2013)}]{2013ApJ...762...96B}
{Bogdanov}, S. 2013, \apj, 762, 96, \dodoi{10.1088/0004-637X/762/2/96}

\bibitem[{{Bogdanov}(2016)}]{2016EPJA...52...37B}
---. 2016, European Physical Journal A, 52, 37,
  \dodoi{10.1140/epja/i2016-16037-x}

\bibitem[{{Bogdanov} \& {Grindlay}(2009)}]{2009ApJ...703.1557B}
{Bogdanov}, S., \& {Grindlay}, J.~E. 2009, \apj, 703, 1557,
  \dodoi{10.1088/0004-637X/703/2/1557}

\bibitem[{{Bogdanov} {et~al.}(2006){Bogdanov}, {Grindlay}, {Heinke}, {Camilo},
  {Freire}, \& {Becker}}]{2006ApJ...646.1104B}
{Bogdanov}, S., {Grindlay}, J.~E., {Heinke}, C.~O., {et~al.} 2006, \apj, 646,
  1104, \dodoi{10.1086/505133}

\bibitem[{{Bogdanov} {et~al.}(2007){Bogdanov}, {Rybicki}, \&
  {Grindlay}}]{2007ApJ...670..668B}
{Bogdanov}, S., {Rybicki}, G.~B., \& {Grindlay}, J.~E. 2007, \apj, 670, 668,
  \dodoi{10.1086/520793}

\bibitem[{{Bruel}(2019)}]{bruel19}
{Bruel}, P. 2019, \aap, 622, A108, \dodoi{10.1051/0004-6361/201834555}

\bibitem[{{de Jager} {et~al.}(1989){de Jager}, {Raubenheimer}, \&
  {Swanepoel}}]{Htest}
{de Jager}, O.~C., {Raubenheimer}, B.~C., \& {Swanepoel}, J.~W.~H. 1989, \aap,
  221, 180

\bibitem[{{Dermer} \& {Sturner}(1994)}]{dermer94}
{Dermer}, C.~D., \& {Sturner}, S.~J. 1994, \apjl, 420, L75,
  \dodoi{10.1086/187166}

\bibitem[{{Dyks} \& {Harding}(2004)}]{DyksHarding2004}
{Dyks}, J., \& {Harding}, A.~K. 2004, \apj, 614, 869, \dodoi{10.1086/423707}

\bibitem[{{Dyks} {et~al.}(2004){Dyks}, {Rudak}, \& {Harding}}]{Dyks2004}
{Dyks}, J., {Rudak}, B., \& {Harding}, A.~K. 2004, \apj, 607, 939,
  \dodoi{10.1086/383587}

\bibitem[{{Foreman-Mackey} {et~al.}(2013){Foreman-Mackey}, {Hogg}, {Lang}, \&
  {Goodman}}]{emcee}
{Foreman-Mackey}, D., {Hogg}, D.~W., {Lang}, D., \& {Goodman}, J. 2013, \pasp,
  125, 306, \dodoi{10.1086/670067}

\bibitem[{{Forestell} {et~al.}(2014){Forestell}, {Heinke}, {Cohn}, {Lugger},
  {Sivakoff}, {Bogdanov}, {Cool}, \& {Anderson}}]{forestell14}
{Forestell}, L.~M., {Heinke}, C.~O., {Cohn}, H.~N., {et~al.} 2014, \mnras, 441,
  757, \dodoi{10.1093/mnras/stu559}

\bibitem[{{Galeazzi} {et~al.}(2014){Galeazzi}, {Chiao}, {Collier}, {Cravens},
  {Koutroumpa}, {Kuntz}, {Lallement}, {Lepri}, {McCammon}, {Morgan}, {Porter},
  {Robertson}, {Snowden}, {Thomas}, {Uprety}, {Ursino}, \&
  {Walsh}}]{galeazzi14}
{Galeazzi}, M., {Chiao}, M., {Collier}, M.~R., {et~al.} 2014, \nat, 512, 171,
  \dodoi{10.1038/nature13525}

\bibitem[{{Gendreau} \& {Arzoumanian}(2017)}]{NICERNatAs}
{Gendreau}, K., \& {Arzoumanian}, Z. 2017, Nature Astronomy, 1, 895,
  \dodoi{10.1038/s41550-017-0301-3}

\bibitem[{{Gendreau} {et~al.}(2016){Gendreau}, {Arzoumanian}, {Adkins},
  {Albert}, {Anders}, {Aylward}, {Baker}, {Balsamo}, {Bamford}, {Benegalrao},
  {Berry}, {Bhalwani}, {Black}, {Blaurock}, {Bronke}, {Brown}, {Budinoff},
  {Cantwell}, {Cazeau}, {Chen}, {Clement}, {Colangelo}, {Coleman},
  {Coopersmith}, {Dehaven}, {Doty}, {Egan}, {Enoto}, {Fan}, {Ferro}, {Foster},
  {Galassi}, {Gallo}, {Green}, {Grosh}, {Ha}, {Hasouneh}, {Heefner}, {Hestnes},
  {Hoge}, {Jacobs}, {J{\o}rgensen}, {Kaiser}, {Kellogg}, {Kenyon}, {Koenecke},
  {Kozon}, {LaMarr}, {Lambertson}, {Larson}, {Lentine}, {Lewis}, {Lilly},
  {Liu}, {Malonis}, {Manthripragada}, {Markwardt}, {Matonak}, {Mcginnis},
  {Miller}, {Mitchell}, {Mitchell}, {Mohammed}, {Monroe}, {Montt de Garcia},
  {Mul{\'e}}, {Nagao}, {Ngo}, {Norris}, {Norwood}, {Novotka}, {Okajima},
  {Olsen}, {Onyeachu}, {Orosco}, {Peterson}, {Pevear}, {Pham}, {Pollard},
  {Pope}, {Powers}, {Powers}, {Price}, {Prigozhin}, {Ramirez}, {Reid},
  {Remillard}, {Rogstad}, {Rosecrans}, {Rowe}, {Sager}, {Sanders}, {Savadkin},
  {Saylor}, {Schaeffer}, {Schweiss}, {Semper}, {Serlemitsos}, {Shackelford},
  {Soong}, {Struebel}, {Vezie}, {Villasenor}, {Winternitz}, {Wofford},
  {Wright}, {Yang}, \& {Yu}}]{2016SPIEGendreau}
{Gendreau}, K.~C., {Arzoumanian}, Z., {Adkins}, P.~W., {et~al.} 2016, in
  \procspie, Vol. 9905, Space Telescopes and Instrumentation 2016: Ultraviolet
  to Gamma Ray, 99051H

\bibitem[{{Guillot} {et~al.}(2016){Guillot}, {Kaspi}, {Archibald}, {Bachetti},
  {Flynn}, {Jankowski}, {Bailes}, {Boggs}, {Christensen}, {Craig}, {Hailey},
  {Harrison}, {Stern}, \& {Zhang}}]{guillot16}
{Guillot}, S., {Kaspi}, V.~M., {Archibald}, R.~F., {et~al.} 2016, \mnras, 463,
  2612, \dodoi{10.1093/mnras/stw2194}

\bibitem[{{Gupta} {et~al.}(2009){Gupta}, {Galeazzi}, {Koutroumpa}, {Smith}, \&
  {Lallement}}]{gupta09}
{Gupta}, A., {Galeazzi}, M., {Koutroumpa}, D., {Smith}, R., \& {Lallement}, R.
  2009, \apj, 707, 644, \dodoi{10.1088/0004-637X/707/1/644}

\bibitem[{{Harding} \& {Muslimov}(2001)}]{harding01}
{Harding}, A.~K., \& {Muslimov}, A.~G. 2001, \apj, 556, 987,
  \dodoi{10.1086/321589}

\bibitem[{{Harding} \& {Muslimov}(2002)}]{harding02}
---. 2002, \apj, 568, 862, \dodoi{10.1086/338985}

\bibitem[{{Jahoda} {et~al.}(2006){Jahoda}, {Markwardt}, {Radeva}, {Rots},
  {Stark}, {Swank}, {Strohmayer}, \& {Zhang}}]{Jahoda06}
{Jahoda}, K., {Markwardt}, C.~B., {Radeva}, Y., {et~al.} 2006, \apjs, 163, 401,
  \dodoi{10.1086/500659}

\bibitem[{{Johnson} {et~al.}(2014){Johnson}, {Venter}, {Harding}, {Guillemot},
  {Smith}, {Kramer}, {{\c{C}}elik}, {den Hartog}, {Ferrara}, {Hou}, {Lande}, \&
  {Ray}}]{Johnson2014}
{Johnson}, T.~J., {Venter}, C., {Harding}, A.~K., {et~al.} 2014, The
  Astrophysical Journal Supplement Series, 213, 6,
  \dodoi{10.1088/0067-0049/213/1/6}

\bibitem[{{LaMarr} {et~al.}(2016){LaMarr}, {Prigozhin}, {Remillard}, {Malonis},
  {Gendreau}, {Arzoumanian}, {Markwardt}, \& {Baumgartner}}]{2016SPIELaMarr}
{LaMarr}, B., {Prigozhin}, G., {Remillard}, R., {et~al.} 2016, in \procspie,
  Vol. 9905, Space Telescopes and Instrumentation 2016: Ultraviolet to Gamma
  Ray, 99054W

\bibitem[{{Lange} {et~al.}(2001){Lange}, {Camilo}, {Wex}, {Kramer}, {Backer},
  {Lyne}, \& {Doroshenko}}]{ELL1}
{Lange}, C., {Camilo}, F., {Wex}, N., {et~al.} 2001, \mnras, 326, 274,
  \dodoi{10.1046/j.1365-8711.2001.04606.x}

\bibitem[{{Lee} {et~al.}(2018){Lee}, {Hui}, {Takata}, {Kong}, {Tam}, \&
  {Cheng}}]{lee18}
{Lee}, J., {Hui}, C.~Y., {Takata}, J., {et~al.} 2018, \apj, 864, 23,
  \dodoi{10.3847/1538-4357/aad284}

\bibitem[{Luo {et~al.}(2019)}]{PINT}
Luo, J., {et~al.} 2019, ApJ

\bibitem[{Mitchell {et~al.}(2018)Mitchell, Winternitz, Hassouneh, Price,
  Semper, Yu, Ray, Wolff, Kerr, Wood, Arzoumanian, Gendreau, Guillemot,
  Cognard, \& Demorest}]{SEXTANTGNC}
Mitchell, J.~W., Winternitz, L.~B., Hassouneh, M.~A., {et~al.} 2018, in Proc.
  41st Annual AAS Guidance \& Control Conference, AAS 18--155

\bibitem[{{Pletsch} \& {Clark}(2015)}]{pc15}
{Pletsch}, H.~J., \& {Clark}, C.~J. 2015, \apj, 807, 18,
  \dodoi{10.1088/0004-637X/807/1/18}

\bibitem[{{Prigozhin} {et~al.}(2016){Prigozhin}, {Gendreau}, {Doty}, {Foster},
  {Remillard}, {Malonis}, {LaMarr}, {Vezie}, {Egan}, {Villasenor},
  {Arzoumanian}, {Baumgartner}, {Scholze}, {Laubis}, {Krumrey}, \&
  {Huber}}]{2016SPIEPrigozhin}
{Prigozhin}, G., {Gendreau}, K., {Doty}, J.~P., {et~al.} 2016, in \procspie,
  Vol. 9905, Space Telescopes and Instrumentation 2016: Ultraviolet to Gamma
  Ray, 99051I

\bibitem[{{Ransom} {et~al.}(2011){Ransom}, {Ray}, {Camilo}, {Roberts}, {{\c
  C}elik}, {Wolff}, {Cheung}, {Kerr}, {Pennucci}, {DeCesar}, {Cognard}, {Lyne},
  {Stappers}, {Freire}, {Grove}, {Abdo}, {Desvignes}, {Donato}, {Ferrara},
  {Gehrels}, {Guillemot}, {Gwon}, {Harding}, {Johnston}, {Keith}, {Kramer},
  {Michelson}, {Parent}, {Saz Parkinson}, {Romani}, {Smith}, {Theureau},
  {Thompson}, {Weltevrede}, {Wood}, \& {Ziegler}}]{rrc+11}
{Ransom}, S.~M., {Ray}, P.~S., {Camilo}, F., {et~al.} 2011, \apjl, 727, L16,
  \dodoi{10.1088/2041-8205/727/1/L16}

\bibitem[{{Ray} {et~al.}(2017){Ray}, {Arzoumanian}, \& {Gendreau}}]{rag17}
{Ray}, P.~S., {Arzoumanian}, Z., \& {Gendreau}, K.~C. 2017, in Proceedings of
  IAU Symposium 337: Pulsar Astrophysics - The Next 50 Years

\bibitem[{{Ray} {et~al.}(2012){Ray}, {Abdo}, {Parent}, {Bhattacharya},
  {Bhattacharyya}, {Camilo}, {Cognard}, {Theureau}, {Ferrara}, {Harding},
  {Thompson}, {Freire}, {Guillemot}, {Gupta}, {Roy}, {Hessels}, {Johnston},
  {Keith}, {Shannon}, {Kerr}, {Michelson}, {Romani}, {Kramer}, {McLaughlin},
  {Ransom}, {Roberts}, {Saz Parkinson}, {Ziegler}, {Smith}, {Stappers},
  {Weltevrede}, \& {Wood}}]{RayPSC}
{Ray}, P.~S., {Abdo}, A.~A., {Parent}, D., {et~al.} 2012, in Third Fermi
  Symposium, eConf C110509 (arXiv:1205.3089)

\bibitem[{{Romani}(1987)}]{1987ApJ...313..718R}
{Romani}, R.~W. 1987, \apj, 313, 718, \dodoi{10.1086/165010}

\bibitem[{{Shklovskii}(1970)}]{1970SvA....13..562S}
{Shklovskii}, I.~S. 1970, \sovast, 13, 562

\bibitem[{{Tauris} \& {van den Heuvel}(2006)}]{tauris06}
{Tauris}, T.~M., \& {van den Heuvel}, E.~P.~J. 2006, {Formation and evolution
  of compact stellar X-ray sources} (Cambridge University Press), 623--665

\bibitem[{{Verner} {et~al.}(1996){Verner}, {Ferland}, {Korista}, \&
  {Yakovlev}}]{verner96}
{Verner}, D.~A., {Ferland}, G.~J., {Korista}, K.~T., \& {Yakovlev}, D.~G. 1996,
  \apj, 465, 487, \dodoi{10.1086/177435}

\bibitem[{{Watts} {et~al.}(2016){Watts}, {Andersson}, {Chakrabarty}, {Feroci},
  {Hebeler}, {Israel}, {Lamb}, {Miller}, {Morsink}, {{\"O}zel}, {Patruno},
  {Poutanen}, {Psaltis}, {Schwenk}, {Steiner}, {Stella}, {Tolos}, \& {van der
  Klis}}]{WattsRMP}
{Watts}, A.~L., {Andersson}, N., {Chakrabarty}, D., {et~al.} 2016, Reviews of
  Modern Physics, 88, 021001, \dodoi{10.1103/RevModPhys.88.021001}

\bibitem[{{Wilms} {et~al.}(2000){Wilms}, {Allen}, \& {McCray}}]{wilms00}
{Wilms}, J., {Allen}, A., \& {McCray}, R. 2000, \apj, 542, 914,
  \dodoi{10.1086/317016}

\bibitem[{{Yao} {et~al.}(2017){Yao}, {Manchester}, \& {Wang}}]{YMW16}
{Yao}, J.~M., {Manchester}, R.~N., \& {Wang}, N. 2017, \apj, 835, 29,
  \dodoi{10.3847/1538-4357/835/1/29}

\bibitem[{{Zavlin}(2006)}]{2006ApJ...638..951Z}
{Zavlin}, V.~E. 2006, \apj, 638, 951, \dodoi{10.1086/449308}

\bibitem[{{Zavlin} {et~al.}(2002){Zavlin}, {Pavlov}, {Sanwal}, {Manchester},
  {Tr{\"u}mper}, {Halpern}, \& {Becker}}]{zavlin02}
{Zavlin}, V.~E., {Pavlov}, G.~G., {Sanwal}, D., {et~al.} 2002, \apj, 569, 894,
  \dodoi{10.1086/339351}

\bibitem[{{Zavlin} {et~al.}(1996){Zavlin}, {Pavlov}, \&
  {Shibanov}}]{1996A&A...315..141Z}
{Zavlin}, V.~E., {Pavlov}, G.~G., \& {Shibanov}, Y.~A. 1996, \aap, 315, 141

\end{thebibliography}



\end{document}